# Three-dimensional disordered conductors in a strong magnetic field: surface states and quantum Hall plateaus.

J.T. Chalker

*Theoretical Physics, University of Oxford, 1 Keble Road, Oxford OX1 3NP, UK*

A. Dohmen

*Institut für Theoretische Physik, Universität zu Köln, Zülpicher Str. 77, 50937 Köln, Germany*

We study localization in layered, three-dimensional conductors in strong magnetic fields. We demonstrate the existence of three phases - insulator, metal and quantized Hall conductor - in the two-dimensional parameter space obtained by varying the Fermi energy and the interlayer coupling strength. Transport in the quantized Hall conductor occurs via extended surface states. These surface states constitute a subsystem at a novel critical point, which we describe using a new, directed network model.

73.40.Hm,72.15.Rn,71.30.+h

The integer quantum Hall effect is one of the most striking phenomena observed in two-dimensional electron systems [1]. Central to it is the existence of phases in which the Hall conductance is constant over a range of values for the Fermi energy. In a quantum Hall phase, electron states at the Fermi energy are Anderson localized within the bulk of a sample, but there exist extended states at the edge of a sample, which are robust against scattering by disorder. It is natural to ask whether quantum Hall phases and edge states are unique to two-dimensional electron systems, or whether they have analogues in three-dimensional conductors.

An obvious way to approach this question is to consider a conductor consisting of layers perpendicular to the applied magnetic field, each of which, in isolation, would exhibit the integer quantum Hall effect. If the inter-layer coupling is weak, it is reasonable to anticipate, with increasing Fermi energy, the sequence of phases sketched in Fig.1: insulator, metal and quantized Hall conductor. Our aim in the following paper is to investigate theoretically this phase diagram and, in particular, the nature of surface states in a three-dimensional quantized Hall conductor.

Studies of three dimensional conductors in quantizing magnetic fields have an extensive history [2]. A variety of situations can be engineered in layered semiconductors. Multi-quantum-well structures with thick barriers between the wells represent the limiting case of uncoupled layers, and simply constitute a number of independent two-dimensional systems in parallel [3]. By contrast, the influence of interlayer coupling is probed in superlattices with appreciable dispersion of the electronic spectrum in the direction perpendicular to the layers. Accurately quantized Hall plateaus are observed [4], as well as an oscillatory variation, with inverse magnetic field, of the transverse and longitudinal diagonal elements of the conductivity tensor, $\sigma_{xx}$ and $\sigma_{zz}$, suggesting an alternating sequence of quantized Hall phases and metallic phases, as in Fig.1. It is possible that some of these features persist in homogeneous semiconductors. In narrow gap semiconductors in the strong magnetic field limit, a temperature independent $\sigma_{xy}$ in conjuction with transverse and longitudinal conductivities, $\sigma_{xx}$ and $\sigma_{zz}$, that both decrease at low temperature [5,6], has been interpreted as an incipient Hall plateau. The foregoing examples are of particular interest in the present context, since it is likely that they show mainly the effect of disorder on single-particle motion. In other settings [2], notably in the spin-density wave phases of Bechgaard salts [7], many-body correlations play an essential role.

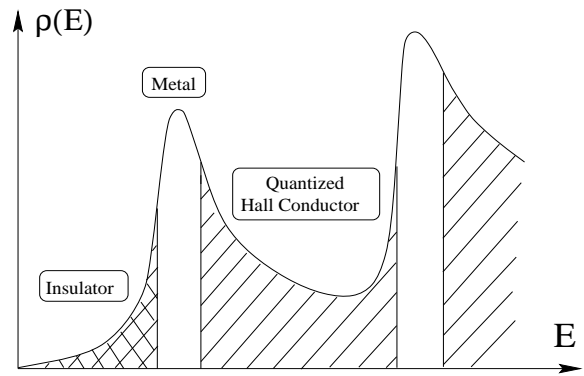

FIG. 1. Schematic phase diagram and density of states, $\rho(E)$, in energy, $E$, for the lowest Landau levels of a layered, three-dimensional conductor in a strong magnetic field.

The theoretical treatment of electrons in a disordered potential and a magnetic field has appealing simplifications in the adiabatic limit, reached if the potential is smooth and the field is strong [8]. These have been exploited for three-dimensional conductors by Azbel [9]. Classically, there are three components to the dynamics, with widely-separated time scales. In the adiabatic limit, the action associated with each component is independently conserved. The shortest timescale is the pe-riod of cyclotron motion about an instantaneous guiding center; oscillations of the guiding center parallel to the magnetic field set the intermediate scale; and at long times the trajectory followed in these oscillations drifts within the plane perpendicular to the field. Quantization of cyclotron and guiding center oscillations leaves a reduced problem, involving only guiding center drift.

Our starting point is a simplified model for the quantum mechanics of this guiding center drift in layered, three-dimensional conductors. It is a natural generalization of the two-dimensional network model for the quantum Hall effect [10]: each layer of the conductor is represented by a copy of the two-dimensional model, and adjacent layers are coupled. Two such coupled layers have been studied previously [11], as a representation of a spin-degenerate Landau level, but the behavior of many layers together has not been investigated before. In detail, the three-dimensional model consists of a network of links, each carrying probability flux in the direction of guiding center drift, which meet at nodes, where flux is scattered between them. Every link is characterized by the phase shift that an electron acquires on transversing it, and randomness is introduced by choosing these phases independently from a uniform distribution. For simplicity, the links are arranged on a regular lattice, as illustrated in Fig.2. Scattering at a node can be specified by a transfer matrix which relates ingoing and outcoming amplitudes $(A_{in}, A_{out})$ on one side of the node, to those $(B_{in}, B_{out})$ on the other. With an appropriate choice of gauge [10]

$$\begin{pmatrix} A_{in} \\ A_{out} \end{pmatrix} = \begin{pmatrix} \cosh(\theta) & \sinh(\theta) \\ \sinh(\theta) & \cosh(\theta) \end{pmatrix} \begin{pmatrix} B_{out} \\ B_{in} \end{pmatrix}. \qquad (1)$$

We choose the scattering parameter, $\theta$, to be the same at all intralayer nodes, with a value, $\theta_1$, related to the Fermi energy, $E$, by [12] $E = \ln(\sinh^2(\theta_1))$. Similarly, we take a second common value, $\theta_2$, at all interlayer nodes, the tunneling amplitude being $t = \tanh(\theta_2)$. The model therefore has a two-dimensional parameter space: $(E, t)$.

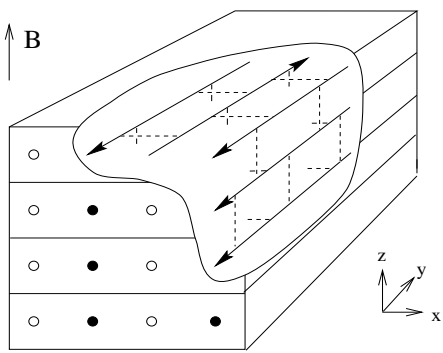

FIG. 2. The 3d network model. Full lines represent links, which carry probability flux in the direction indicated by the arrows. Dashed lines represent nodes. Ingoing (outcoming) links on the x-z face are denoted by ∘ (•).

We have investigated the phase diagram of this model numerically, using standard transfer matrix techniques to calculate the localization length in quasi-one dimensional samples, and a finite-size scaling analysis to extract the bulk behavior [13]. We study systems of cross-section $M \times M$ and length $L$ for $M \leq 12$ and $L \leq 8.10^4$, obtaining Lyapunov exponents with statistical error $\leq 1.5\%$. To concentrate initially on the properties of bulk states, we apply periodic boundary conditions in the directions transverse to $L$. As a simple test of our approach, we have checked that we obtain the same phase diagram from calculations with the layers of the model arranged either parallel or perpendicular to $L$.

The results are displayed in Fig.3. Without interlayer coupling ($t = 0$) we reproduce properties of the two-dimensional model: extended states exist only at the Landau level center ($E = 0$). Non-zero coupling ($t > 0$) gives rise to a band of extended states, having a width in energy, $W(t)$, that increases with $t$. For $t \ll 1$, one expects [14] $W(t) \propto t^{1/\nu_{2d}}$, where $\nu_{2d}$ is the critical exponent for the divergence of the localization length, $\xi_{2d}$, in an isolated layer, and this form is consistent with our data. Bulk states in both the low- and high-energy tails of the Landau level remain localized even for the largest interlayer coupling investigated. We note that, because the network model omits inter-Landau level scattering, it cannot capture behaviour in the strong disorder or weak magnetic field limits, when extended states presumably levitate in energy [15].

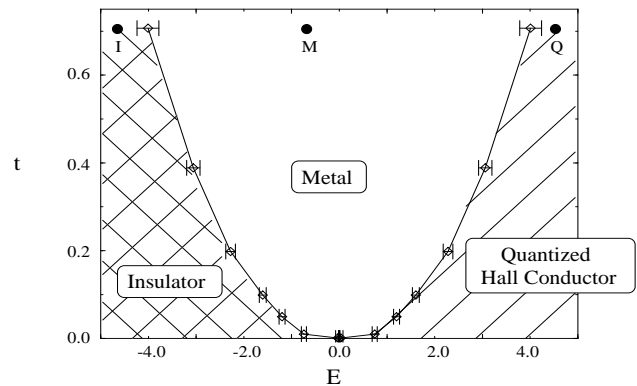

FIG. 3. Phase diagram of the 3d network model. Data obtained at the points marked $I$, $M$ and $Q$ are shown in Fig.4.

In order to investigate surface states in the model, we compare the spectra of Lyapunov exponents in samples with periodic and hard-wall boundary conditions. One expects surface states, present only in the quantized Hall conductor with hard-wall boundary conditions, to carry current without backscattering in samples with large cross-section ($M \gg 1$), and hence to be associated with vanishing Lyapunov exponents. We study quasi-one-dimensional samples with (using axes defined in Fig.2) their long side parallel to $y$, periodic boundary conditions in the $z$-direction, and either periodic or hard-wall boundary conditions in the $x$-direction. With periodic boundary conditions, the Lyapunov exponent spectrum is gapless in the metal, and has a gap (of size $\xi_B^{-1}$ for large $M$, where $\xi_B$ is the bulk localization length) in both the insulator and the quantized Hall conductor. There are only small changes in the spectra for the insulator and metal on switching to hard-wall boundary conditions, and we attribute these changes to finite-size effects. By contrast, in the quantized Hall conductor this switch has a dramatic influence on the distribution of Lyapunov exponents, shown in Fig.4. With hard-wall boundary conditions, $M$ of the (positive) exponents are small, and decrease with increasing $M$, while the values of the others are little altered, indicating that there are $M$ surface states. This interpretation is reinforced by examining the eigenvectors of the transfer matrix corresponding to the $M$ smallest Lyapunov exponents (Fig.4, inset): their amplitude is concentrated overwhelmingly near the sample surface.

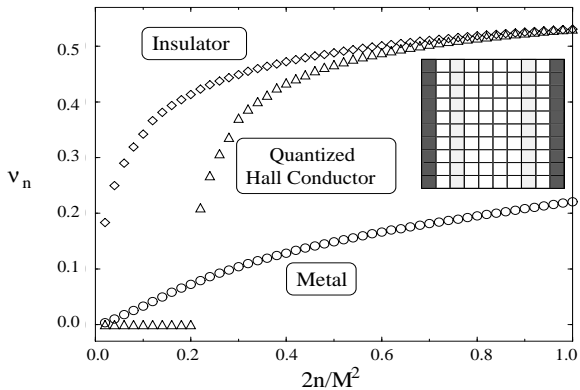

FIG. 4. Spectra of positive Lyapunov exponents, $\nu_n, n = 1 \ldots M^2/2$, at three points in the phase diagram, marked I ($\diamond$), M ($\circ$) and Q ($\triangle$) in Fig.3, for sample size $M = 10$. Inset: Amplitude distribution for the transfer matrix eigenvector with the smallest positive Lyapunov exponent, in the quantized Hall conductor. Squares represent links of the model and are shaded according to the mean probability flux, $p$ carried by that link: black, $p > 0.04$; grey, $0.04 \geq p > 2.10^{-4}$; white, $2.10^{-4} > p$.

Next, we focus on the subsystem of surface states by considering the high energy tail of the Landau level, where the bulk localization length within our model is very short. Then each layer supports an edge state, which is decoupled from localized bulk states but coupled to edge states in adjacent layers. Edge states in different layers carry probability flux in the same sense, and we can represent the surface using the two-dimensional *directed* network model illustrated in Fig.5, which incorporates randomness via link phases, as in the three-dimensional model, and has a single parameter, the tunneling amplitude between layers, $t = \tanh(\theta_2)$. This model is clearly highly anisotropic, and represents an example of directed scattering, studied previously in other contexts [16]. Consider a finite sample in the form of a cylinder, of circumference $C$ and height $L$, with its axis parallel to the magnetic field. Charge transport around the circumference is rather simple. It is characterized by the Hall conductance, and the restriction of scattering to the forward direction ensures that this is quantized. Transport in the direction parallel to the magnetic field is more subtle. We find that the system is at a critical point, and the average conductance, $g_{zz}$, depends on the aspect ratio, $L/C$. For $L/C \ll 1$, the surface states have a finite conductance per square, $\sigma_\square$, which is a function of the tunneling amplitude, $t$, and $g_{zz} = (C/L)\sigma_\square$. Proportionality of $g_{zz}$ to sample circumference, $C$, rather than cross-sectional area, is, of course, a signature of conduction by the surface, rather than the bulk, and is a characteristic of the quantized Hall conductor. In the opposite limit, $L/C \gg 1$, the system is quasi-one-dimensional, states have a finite localization length, $\xi$, along the cylinder axis, and $g_{zz}$ decreases exponentially with $L$. The system is revealed to be critical by the fact that $\xi \propto C$, with an amplitude ratio $A \equiv \xi/C$. Remarkably, it is possible to calculate both $\sigma_\square$ and $A$ analytically.

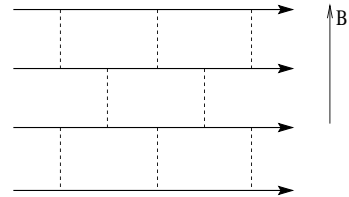

FIG. 5. The directed network model. Full and dashed lines represent links and nodes, as in Fig.2.

To evaluate $\sigma_\square$, we express each element, $t_{ij}$, of the $C \times C$ transmission matrix, $\boldsymbol{t}$, between ends of the cylinder in terms of a sum over Feynman paths. Because of the directed character of the model, self-intersecting paths must wind at least once around the cylinder, and make no contribution to the sum in the limit $L \ll C$. Retaining only those paths that do not self-intersect, it is straightforward to average $|t_{ij}|^2$ over the link phases. From this, using the Landauer-Buttiker formula [17], we find

$$\sigma_\square = (e^2/h)t^2/(1-t^2). \quad (2)$$

To obtain the amplitude ratio, $A$, we first parameterize the eigenvalues of $\boldsymbol{t}^\dagger \boldsymbol{t}$ as $\cosh^{-2}(\nu_n L/C)$, with $\nu_1 \leq \nu_2 \leq \ldots \leq \nu_C$. In the limit $L \gg C$, the $\{\nu_n\}$ are proportional to Lyapunov exponents, and therefore self-averaging; $A = \nu_1^{-1}$. Moreover, one expects [18] for $n \ll C$ that $\nu_n = n \cdot \nu_1$. In the converse limit, $C \gg L$, rigidity in the spectrum of $\boldsymbol{t}^\dagger \boldsymbol{t}$ suppresses fluctuations in the $\{\nu_n\}$, and one again expects [18] that $\nu_n = n \cdot \nu_1$. In this case

$$\sigma_\square = \lim_{L\to\infty} \lim_{C\to\infty} \frac{L}{C}\frac{e^2}{h} \sum_{n=1}^{C} \cosh^{-2}(Ln\nu_1/C) = \frac{e^2}{h}\nu_1^{-1}.$$

Making the conjecture that the value of $\nu_1$ is the same in both limits, we find

$$A = t^2/(1-t^2). \qquad (3)$$

In order check this result, we have evaluated the amplitude $A$ numerically, using cylinders of size $C \leq 64$ and $L \leq 10^6$, and extrapolating $\xi/C$ to large $C$. The coincidence shown in Fig.6 between the numerical data and Eq.(3) is striking.

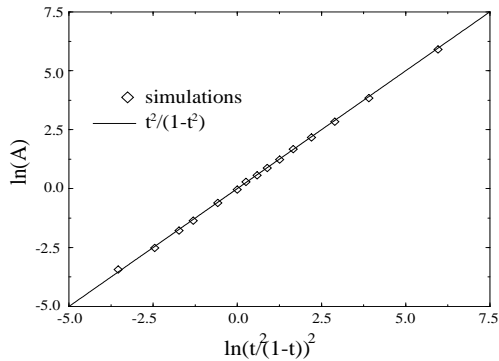

FIG. 6. Comparison of the results of a numerical calculation of the amplitude $A$ ($\diamond$) with the analytic expression (—)

Finally, we return to the three-dimensional model and consider bulk critical phenomena at the transitions from the metal to the insulator or quantum Hall conductor. Since bulk states are localized in both the latter phases, we expect the same critical behavior at each transition; within the network model, this is guaranteed by an exact symmetry, arising because higher Landau levels are omitted. We calculate the localization length exponent, $\nu$, for comparison with earlier results, obtained from other models of the metal-insulator transition in three dimensions and strong magnetic field. We find, using a standard analysis [13], $\nu = 1.45 \pm 0.25$, which is consistent with the value $\nu = 1.35\pm0.15$, obtained both for a layered system and a tight-binding model. [19,20].

In summary, we have introduced a model of a layered conductor in a magnetic field and shown by numerical simulation that it has the phase diagram of Fig. 1. We have identified the surface states of a three-dimensional quantized Hall conductor as a critical system, described by a second model, the directed network model, for which key quantities are calculable analytically.

We are grateful for discussions with M. Janssen, D.E Khmelnitskii and B. Shapiro. This work was supported in part by SERC grant GR/GO 2727, EC grant SCC CT90 0020 and the Sonderforschungsbereich 341.

[1] For reviews, see *The Quantum Hall Effect*, edited by R.E. Prange and S.M. Girvin (Springer, Berlin, 1990); and M. Janßen, O. Viehweger, U. Fastenrath and J. Hajdu *Introduction to the Theory of the Integer Quantum Hall Effect* (VCH, Weinheim, 1994).
[2] For a review, see: B.I. Halperin, Jpn. J. Appl. Phys. **26** Suppl.26-3, 1913 (1987).
[3] T. Haavasoja *et al*, Surf. Sci. **142** 294 (1984).
[4] H.L. Störmer *et al*, Phys. Rev. Lett. **56** 85 (1986).
[5] S.S.Murzin, Pis'ma Zh. Eksp. Teor. Fiz. **44**, 45 (1986) [JETP Lett. **44**, 56 (1986)]; U. Zeitler *et al*, J.Phys. Cond. Matt. **6**, 4289 (1994).
[6] R.G. Mani, Phys. Rev. B**41**, 7922 (1989).
[7] D. Poilblanc *et al*, Phys. Rev. Lett. **58** 270 (1986).
[8] T.G. Northrup, *The adiabatic motion of charged particles* (Interscience, New York, 1963)
[9] M.Y. Azbel, Solid State Commun. **54**, 127 (1985); M.Y. Azbel and O. Entin-Wohlman, Phys. Rev. B**32** 562 (1985).
[10] J.T.Chalker, P.D. Coddington, J. Phys. C **21**, 2665 (1988).
[11] D.K.K. Lee and J.T. Chalker, Phys. Rev. Lett. **72**, 1510 (1994); Z.Q. Wang, D-H Lee, and X-G Wen, Phys. Rev. Lett. **72**, 2454 (1994).
[12] H.A. Fertig and B.I. Halperin, Phys. Rev. B **36**, 7969 (1987).
[13] J.L. Pichard and G. Sarma, J. Phys C **17**, 4111 (1981); A. MacKinnon and B. Kramer, Phys. Rev. Lett. **47**, 1546 (1981); Z. Phys. B**53**, 1 (1983).
[14] Consider the interlayer coupling, $\hat{t}$, as a perturbation on localized eigenstates, $\psi$ and $\psi'$ in isolated layers. Let $\Delta$ be the level spacing in a layer of area $\xi_{2d}^2$. At the mobility edge, $|\langle\psi|\hat{t}|\psi'\rangle| \sim \Delta$, from the Thouless criterion. We expect $|\langle\psi|\hat{t}|\psi'\rangle| \sim t/\xi_{2d}$ and hence $t \sim \xi_{2d}^{-1}$ or $|E| \sim t^{1/\nu_{2d}}$ at the transition. A similar argument is given in Ref [19], but with $|\langle\psi|\hat{t}|\psi'\rangle| \sim t$, which leads to $|E| \sim t^{1/2\nu_{2d}}$, in conflict with our data.
[15] D.E. Khmelnitskii, Phys. Lett. **106A**, 182 (1984); R.B. Laughlin, Phys. Rev. Lett. **52**, 2304 (1984).
[16] L. Saul, M. Kardar, and N. Read, Phys. Rev. A**45**, 8859 (1992); C. Barnes, B.L. Johnson and G. Kirczenow, Phys. Rev. Lett. **70**, 1159 (1993).
[17] M. Büttiker *et al*, Phys. Rev. B**31**, 6207 (1985).
[18] J.L. Pichard and G. André, Europhys. Lett. **2**, 477 (1986); A.D. Stone *et al* in *Mesoscopic Phenomena in Solids*, edited by B.L. Altshuler, P.A. Lee and R.A. Webb (North-Holland, Amsterdam, 1991).
[19] T. Ohtsuki, B. Kramer, and Y. Ono, J. Phys. Soc. Jpn. **62** 224 (1993).
[20] M.Hennecke, B.Kramer and T. Ohtsuki, Europhys. Lett. **27**, 389 (1994).